\documentclass{JINST}

\title{A new instrument for high statistics measurement of photomultiplier characteristics}

\author{C. Bozza$^a$, T. Chiarusi$^b$, M. Costa$^c$, F. Di Capua$^{d,e}$, V. Kulikovskiy$^c$\thanks{Corresponding
author.}, R. Mele$^{d,e}$, P.~Migliozzi$^e$, C. M. Mollo$^e$\thanks{Corresponding
author.}, C. Pellegrino$^b$, G. Riccobene$^c$~ and D. Vivolo $^{d,e}$\\
\llap{$^a$} University of Salerno and INFN Gruppo Collegato di Salerno,\\
Via Giovanni Paolo II 132, Fisciano 84084, Italy\\
\llap{$^b$}INFN Sezione di Bologna,\\
v.le C. Berti-Pichat, 6/2, Bologna 40127, Italy\\
\llap{$^c$}Laboratori Nazionali del Sud,\\
  via Santa Sofia 62, Catania 95123, Italy\\
\llap{$^d$} Universit\'a "Federico II" di Napoli,\\
  Complesso Universitario di Monte S. Angelo via Cintia ed. 6, 80126 Napoli,Italy\\
\llap{$^e$}INFN Sezione di Napoli,\\
  Complesso Universitario di Monte S. Angelo via Cintia ed. 6, 80126 Napoli,Italy\\
  E-mails: \email{kulikovs@ge.infn.it}, \email{maximil@na.infn.it}}

\abstract{Since the early days of experimental particle physics photomultipliers (PMTs) have played an important role in the detector design. Thanks to their capability of fast photon counting, PMTs are extensively used in the new-generation of astroparticle physics experiments, such as air, ice and water Cherenkov detectors. Small size PMTs ($\leq $ 3 inches diameter) show little sensitivity to the Earth magnetic field, small transit time, stable transit time spread; the price per photocathode area is less comparing to the one for the large area PMTs, typically used so far in such applications. Together with developments and reduced price of multichannel electronics, the use of PMTs of 3-inches or smaller diameter is a promising option even for nowadays large volume detectors. 

In this paper we report on the design and performance of a new instrument for mass characterisation of PMTs (from 1 inch to 3 inches size), capable to calibrate hundreds of PMTs  per day and provide measurements of dark counts, signal amplitude, late-, delayed-, pre- and after-pulses, transit time and transit time spread.}


\begin{document}

\section{Introduction}
The search for high-energy particles, such as high-energy neutrinos, cosmic-rays and gammas, pushed the astroparticle physics community to increase the dimensions of experimental apparatuses to enormous sizes and consequently to use thousands of particle detectors.
Most of the experimental apparatuses under construction, or proposed, make use of photon detection as direct or indirect signature for particle identification.
Despite their 80 year-long history, PMTs are still considered optimal detectors for photon detection thanks to their excellent timing and single photon counting capability.

The challenge for building detectors based on vast number of PMTs is then twofold: to design read-out electronics capable to interface the desired number of channels and to set-up proper instruments to qualify and calibrate the large number of PMTs to be used in the detector.

In order to significantly speed up the massive test and calibration of a huge number of PMTs, we developed an instrument (from now on dubbed DarkBox) capable to tune the high voltage (HV) to have equalised gain, and to measure timing resolution, charge resolution, dark counts and spurious pulse rates.

Several challenges were accepted for the DarkBox realization:
\begin{itemize}
\item optically tight box to keep the laser and spurious photons detection rates well below the PMT dark current rates;
\item  mechanical design allowing fast PMTs loading and unloading;
\item electronics capable of tuning each PMT independently;
\item external light source that illuminates each PMT with single photons; the photons arrival time to each PMT photocathode area is known with sub-ns precision;
\item data acquisition system with sub-ns time stamping accuracy and precision;
\item automatisation of the PMT qualification procedures.
\end{itemize}

The DarkBox was initially designed to provide a facility for fast automatic test of 3-inch PMTs chosen by the KM3NeT Collaboration for the construction of an underwater neutrino telescope in the Mediterranean Sea~\cite{KM3TDR, PPM_DOM, PPM_DU}.
For this purpose the DarkBox data acquisition system was designed in compliance with requests from the KM3NeT Collaboration~\cite{DAQ, CU}. The current mechanical design allows the simultaneous test of sixty-two 3-inch PMTs. The design is versatile and can be adapted to different PMT types with minimal changes.

In this work we will describe the DarkBox setup (Section~\ref{s:setup}) including its mechanics, optical and electronic components. We will also describe the complete PMT test procedure and the software used (Section~\ref{s:analysis}) and we will present the results of the qualification of this new instrument (Section~\ref{s:qualification}).

\section{DarkBox setup description}
\label{s:setup}
\begin{figure}
\center
\includegraphics[width=0.9\textwidth]{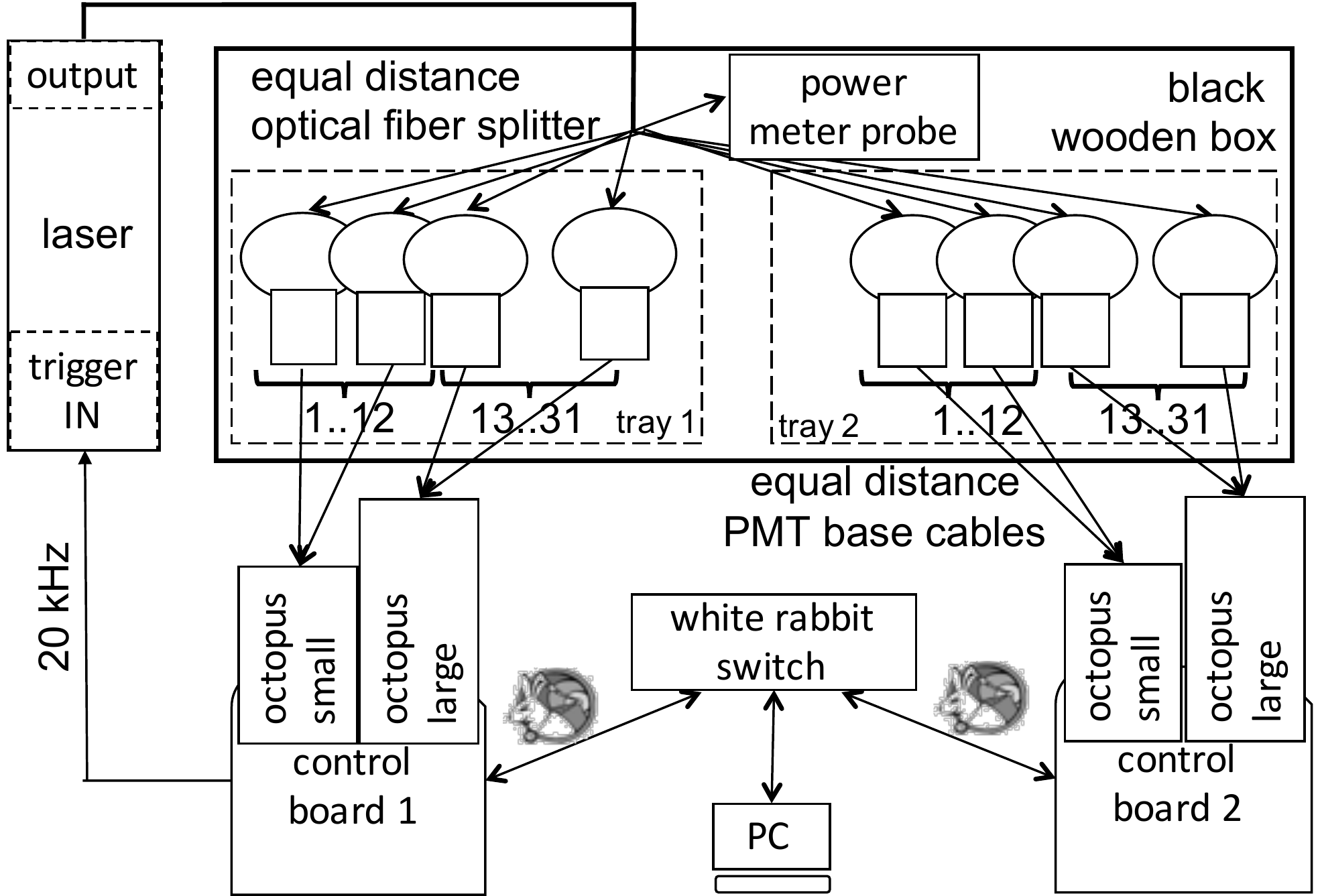}
\caption{DarkBox experimental setup scheme.}
\label{fig:scheme}
\end{figure}

The DarkBox setup is schematically shown in Figure~\ref{fig:scheme}. It consists of a wooden box and removable trays designed to host PMTs under test. A time-calibrated electrical cabling system was realised to connect PMTs to the data acquisition system placed outside the box, maintaining for all PMTs the same time signal delay. A picosecond accuracy laser and a calibrated optical splitting system are used to distribute single photon signal to all PMTs. For the particular application of the KM3NeT PMTs tests the front-end electronics, data management and analysis software are based on the technologies developed by the KM3NeT Collaboration~\cite{CLB}.

\subsection{Mechanics}
\begin{figure}
\center
\includegraphics[width=0.9\textwidth]{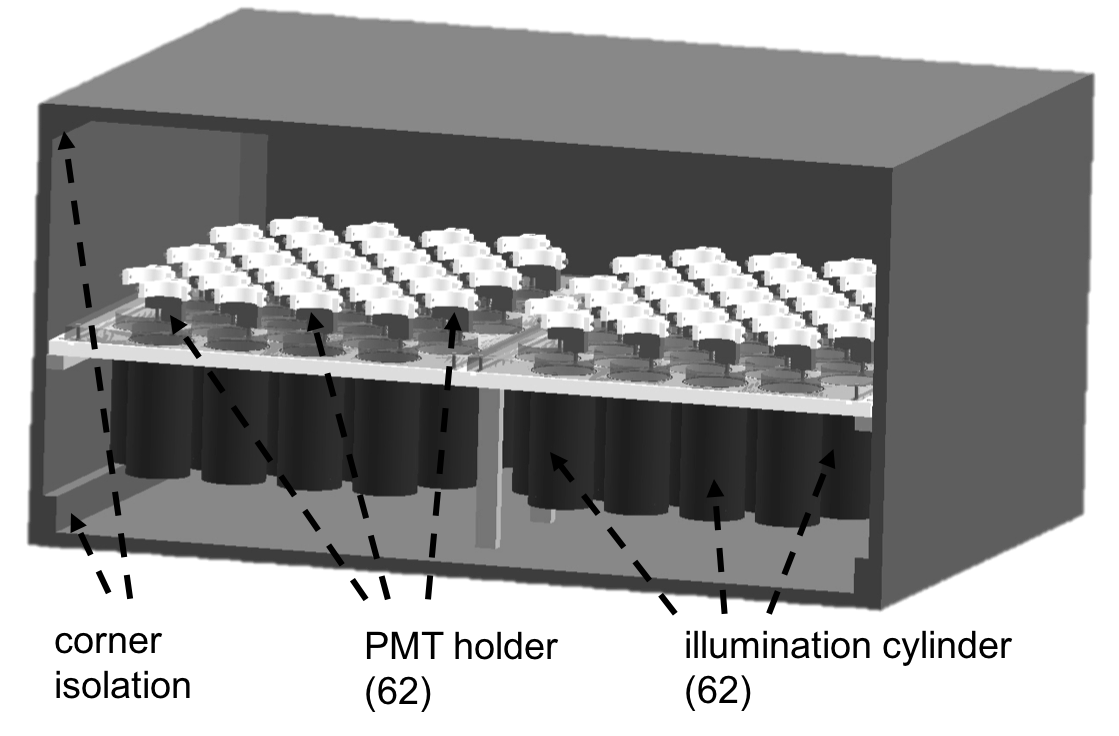}
\caption{Wooden dark box with two removable PMT holder trays placed above the support structure. The latter has plastic cylinders to provide uniform PMT illumination and optically isolate each test socket. Wooden bars on the inner corners and edges are added to ensure the light-tightness.}
\label{fig:mechanical2}
\end{figure}

{\bf Wooden box.} The black wooden box where the PMTs are qualified has the following dimensions: 120~cm x 88~cm x 58~cm (height) to allow simultaneous characterisation of sixty-two 3-inch PMTs. Its walls are black painted externally and internally. As can be seen in Figure~\ref{fig:mechanical2}, particular attention has been devoted to the edges and the corners, usually the most critical regions, to guarantee light tightness. A vertically opening front door (not shown in the picture) is locked tightly by two lever-operated latches on each side (left and right).

{\bf Trays.} Removable trays, loadable before the insertion into the DarkBox, allow easy and fast PMTs reinstallation, thus optimising the operation time. During the measurement the box is filled with two trays that can accommodate 31 PMTs each. Metal arms are attached to each tray on both sides for the grip during the trays exchange operation.

\begin{figure}
\center
\includegraphics[height=.43\textwidth]{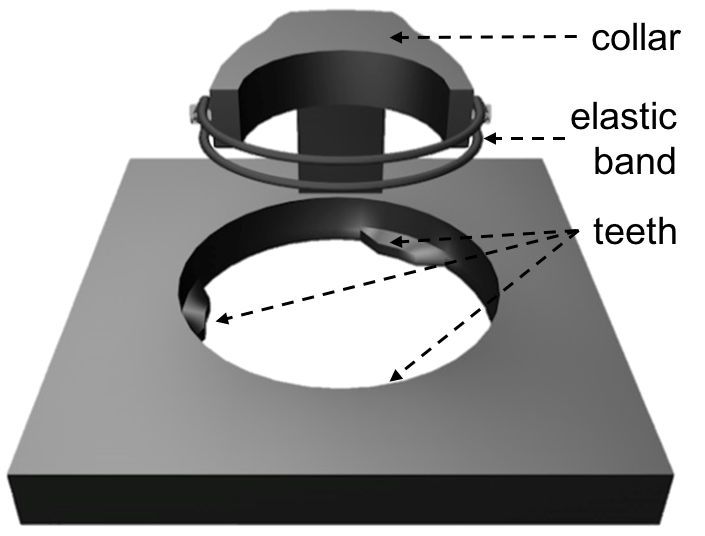}
\includegraphics[height=.48\textwidth]{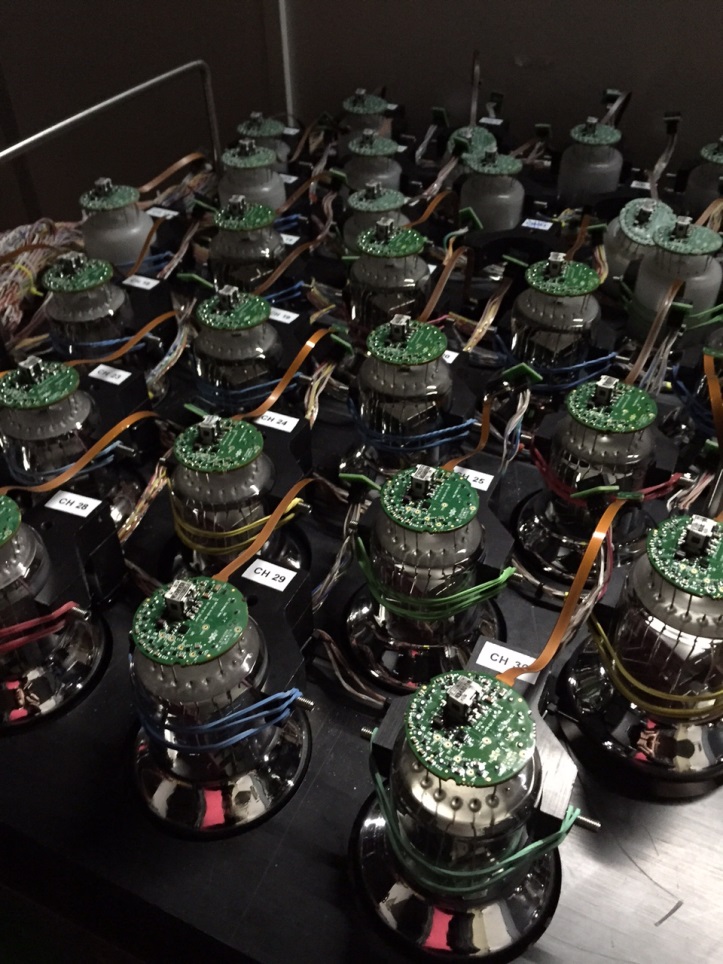}
\caption{The DarkBox tray holder unit (left) and the tray loaded with 31 PMTs (right).}
\label{fig:holderref}
\end{figure}

{\bf PMT holders.} The photomultipliers are maintained in a vertical position with photocathode area looking down. Polyvinyl chloride collars and elastic bands gently fix PMTs in the area around the dynode chain (Figure~\ref{fig:holderref}). Using this solution the photocathode is not in contact with the structure. This is mandatory for PMTs fed with negative voltage at the cathode. Indeed, the proximity of the photocathode area to the mechanical structures may produce electric discharges, dramatically increasing the PMT noise~\cite{Flyckt}. The holders have tiny teeth on the lower openings (Figure~\ref{fig:holderref}, left) to prevent the PMTs from falling off during loading and unloading operations. 

{\bf Support flatbed.} The trays are placed on a flatbed support structure made of polyvinyl chloride. The upper surface of the support structure and the tray bottom are flat which allow heavy tray to be gently slid in. The plate of the support structure has 62 circular holes with dark cylinders attached to their bottom (Figure~\ref{fig:mechanical2}). The cylinders are used for the PMT illumination described in the following section.

\subsection{Laser calibration system}
The measurement of the PMT gain, transit time and spurious signals requires homogeneous PMT photocathode illumination with single photons. The arrival time of the photons to the PMT surface should be known with sub-nanosecond precision for the timing characteristics measurements. A single short-pulse laser with a fiber optic system to deliver the photons to each PMT was chosen for the operation simplicity and the cost. The accuracy and precision of the system is below nanosecond, including pulse width, start time accuracy and jitter. 

{\bf Laser.} A laser controller\footnote{Pilas EIG2000DX, http://www.onefive.com/pilas.html} equipped with a 470~nm laser head is used in external trigger input mode and the trigger is generated by the data acquisition system to provide common time reference for the emitted photons and the detected PMT signal (Section~\ref{s:DAQ}). 
The jitter between the synchronization trigger signal and the optical signal is typically smaller than 50~ps with an optical signal pulse width of about 30~ps (FWHM). 

\begin{figure}
\center
\includegraphics[width=0.8\textwidth]{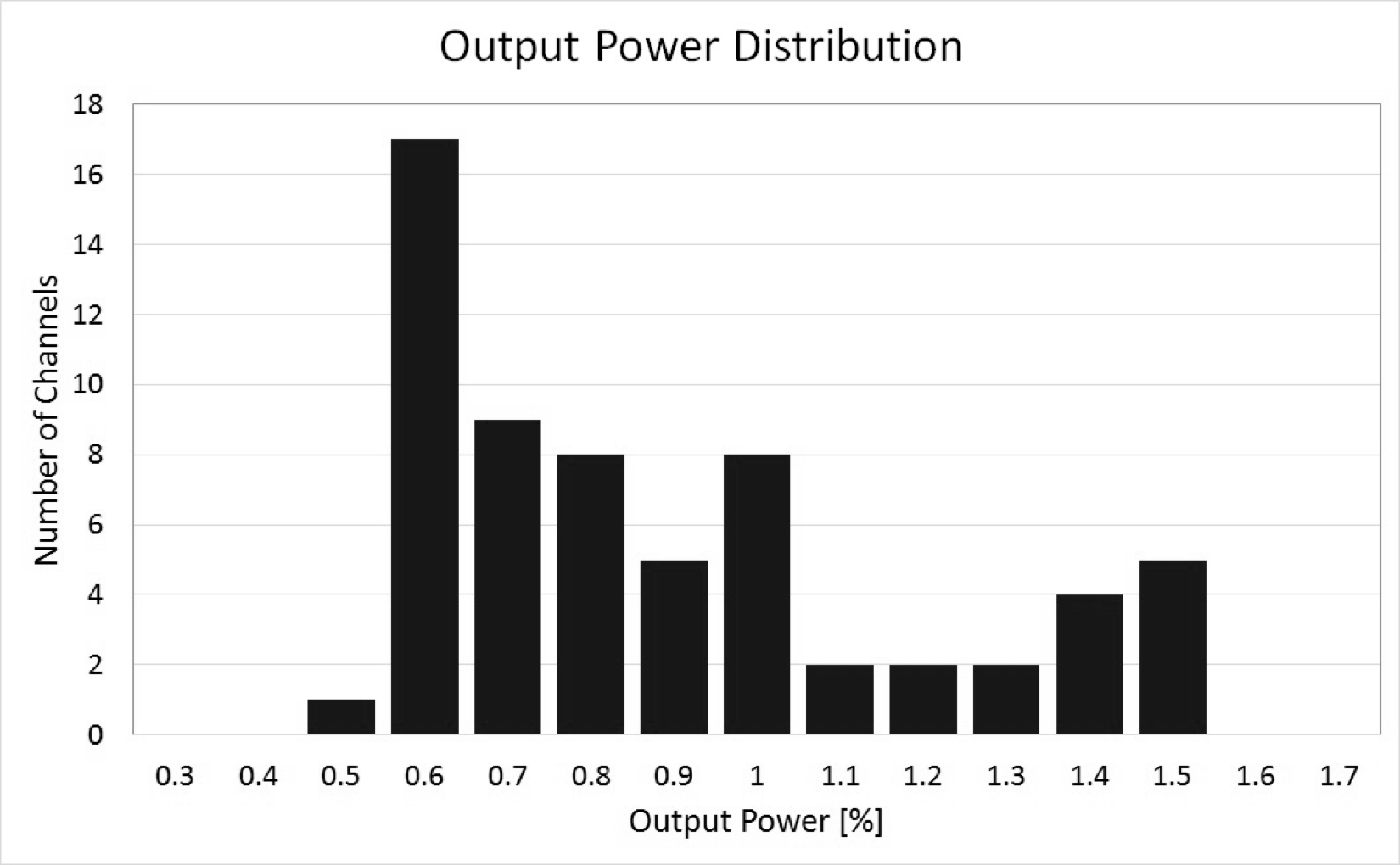}
\caption{The splitter output power distribution for each channel.}
\label{fig:splitter}
\end{figure}

{\bf Optical splitter.} In order to guarantee that photons arrive with the same delay and quantity at each PMT, we used an optical splitter with good uniformity in power distribution and equal light paths for each channel. It is realised with an input fibre having 600~$\mu$m core diameter and 70 output channels, coupled with the main fiber, having 25~$\mu$m diameter fiber core each. The output power distribution for each channel is shown in Figure~\ref{fig:splitter}. The uniformity is such that all outputs have a fraction of the input power in the range 0.5\% - 1.5\%. The lengths of the output fibers are made equal with millimeter precision to equalise the time delays.

\begin{figure}
\center
\includegraphics[width=.9\textwidth]{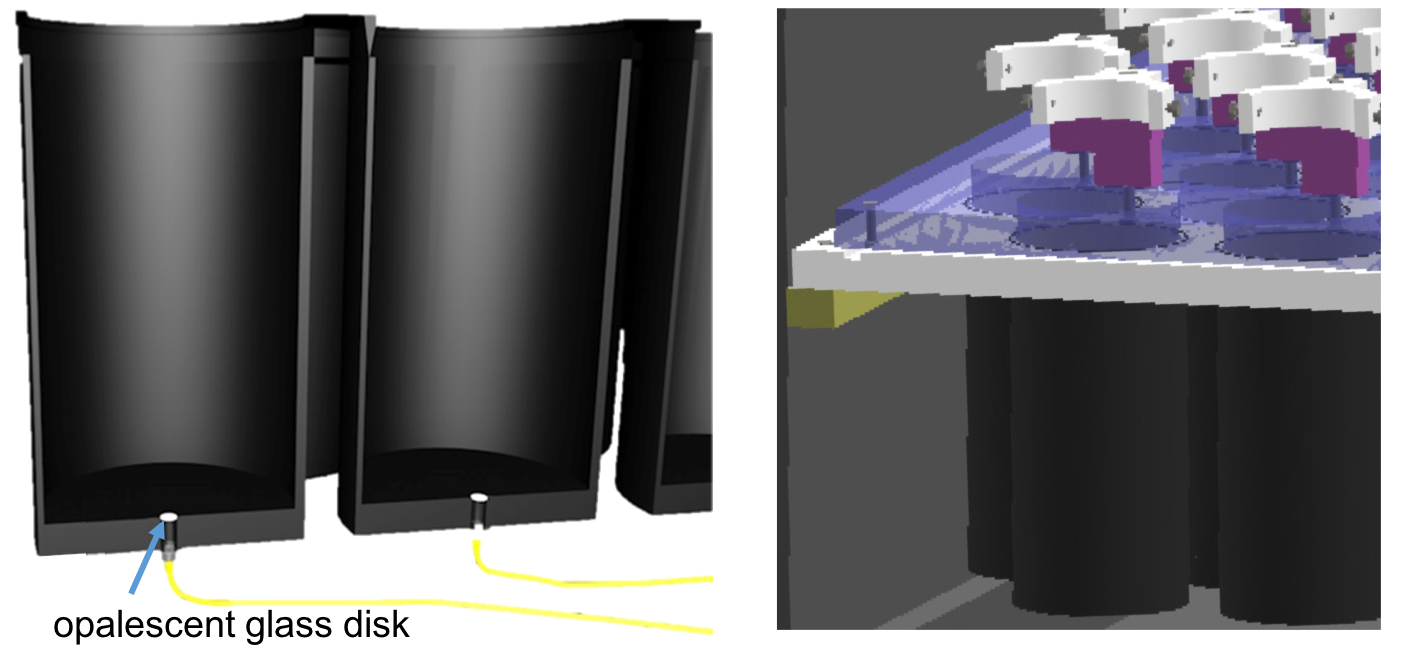}
\caption{The inner structure of the dark cylinders (left) and the external view of the cylinder with a PMT holder (right).}
\label{fig:cylinder}
\end{figure}

{\bf Uniform illumination setup.} In working environment PMTs are illuminated uniformly. Therefore qualification tests should reproduce similar conditions. The uniform illumination of the photocathode is obtained by placing PMT upside-down on top of a black cylinder (Figure~\ref{fig:cylinder}) which has a small central hole in the bottom for the entrance of the optical fiber. An opalescent diffusing glass disk is coupled with the optical fiber and placed in the hole. This helps to reach better illumination uniformity (near lambertian distribution of the photons on the PMTs photocathode). The cylinders are used to provide space for the light cone expansion and to optically isolate each single PMT from the other sockets.

{\bf Calibration: single photoelectron mode.} A tunable (passive) optical attenuator is placed in the optical fiber distribution system to reach the single photon operation mode. The attenuator has a dynamic range of about 20~dB.
In case more optical power is needed the attenuator can be adjusted without changing the laser setting. A spare fiber of the splitting system is used as a reference and monitored to check the single photoelectron operation mode by connecting it to a power meter\footnote{Newport model 2936-C (equipped with a probe Newport mod. 918D-SL-OD3)}, whose sensitivity is 0.4~pW. To monitor the laser power, the laser frequency is increased to reach the power meter sensitivity while keeping stable the emission power itself. 

{\bf Calibration: optical system latency.} The delay between the electric trigger signal and the optical pulse signal arriving to the end of the splitter fibers was measured using an oscilloscope\footnote{Le Croy, Wavescanner Z610X} coupled with optical to electrical converter\footnote{LeCroy OE415}. Propagation time of the light in the cylinder is calculated including the time spread due to cylinder height and PMT radius. The full time delay between the trigger generation and photons arrival to the PMT surface is evaluated with an accuracy of about 100~ps.

\section{Case study: calibration of KM3NeT PMTs}
In order to measure the characteristics of the PMTs used by the KM3NeT Collaboration, a dedicated data acquisition system was set up and operated.

\subsection{Front-end electronics and data acquisition system}
\label{s:DAQ}
{\bf PMT assembly, PMT base.}
The KM3NeT PMT assembly is composed by a 3-inch PMT~\footnote{Hamamatsu  R12199} coupled with a custom active base, that provides the needed high voltage by means of a Cockcroft-Walton generator~\cite{PMT_base}. The base has a control chip with a serial communication interface to set up the high voltage and to provide a unique identifier.

The base electronics shapes the anode signal and converts it into a low voltage differential signal whose time duration is proportional to the signal charge. The discriminator threshold is adjustable through the PMT base control chip; the default value used corresponds to 0.3~photoelectrons (p.e.).

{\bf Control board with octopus boards.} For each group of 31 PMTs the control and data acquisition is performed by using a custom FPGA-based control board, that is designed and used in the KM3NeT optical modules. The interface between each PMT base and the control board is obtained through two so-called octopus boards of different sizes: a small octopus board (12 channels) and a large one (19 channels). 

Time-delay equalised cables are used to connect the PMT assemblies and the control boards. A time-to-digital converter in the control boards digitizes the PMT differential signals, recording the start pulse time and time over threshold (ToT) that is proportional to the PMT charge. Each control board packs the incoming data from the 31 PMTs into ethernet data packets and sends them through a local ethernet network. The management of the control boards is performed using a data acquisition PC through the same ethernet connection.
 
{\bf Data acquisition setup.} The data acquisition setup consists of a Personal Computer (PC) connected through copper ethernet cable to a White Rabbit Switch\footnote{http://www.ohwr.org/projects/white-rabbit/wiki}. Thanks to the White Rabbit gear installed on the FPGA of the switch (White Rabbit Master) and on the two control boards (White Rabbit Slaves), the two control boards are kept synthonised (same clock frequency) and synchronised (same phase delay) with sub-nanosecond accuracy. The connection between the switch and the control boards is performed via single optical fiber lines following the White Rabbit standards. The PMT data and management commands are transmitted between the  control boards and the PC through the switch following the same connection lines.

The trigger input signal for the laser controller is generated by one of the control boards. Therefore, it is perfectly synchronised with the data acquisition electronics. The trigger frequency can be remotely adjusted in the range 0.1 -- 100~kHz.

\subsection{Readout, acquisition and analysis software}
\label{s:analysis}
For this application we developed a custom manager tool based on Java that is interfaced with the KM3NeT software tools to read-out PMT data from the control boards and upload/download recorded parameters into a relational database.

The DarkBox manager performs the following operations:
\begin{itemize}
\item define a unique identification tag for each test; create a directory where data and information for each test are stored; 
\item request identifiers of the connected PMT (they are provided by the PMT base control chips);
\item interface with the KM3NeT database; using the PMT identifiers recover from the database the default PMT settings;
\item interface with KM3NeT detector control and data acquisition software; set proper threshold and HV for each PMT; set test run numbers (to identify different measurement phases); start and stop run; record raw PMT data;
\item create ROOT\footnote{https://root.cern.ch} files for data analysis; analyse ROOT files to measure the required PMT characteristics;
\item produce XML and PDF files with the results to be uploaded into the database. 
\end{itemize}

\subsection{PMT test procedures}
The characterisation of each PMT consists of three steps: determination of the operational HV to reach equalised gain (high voltage tuning procedure), measurement of dark counts, measurement of PMT time characteristics and of spurious pulses. The whole procedure takes about 10 hours to be completed. The largest fraction of this time is devoted to the darkening, i.e. the powered PMTs are left in the dark to recover from the light exposure.

\subsubsection{High voltage tuning}
In the KM3NeT detector the PMT impulse charge is measured  with ToT technique.
In KM3NeT the gain of all PMTs is equalised to a reference value of $3\times10^6$. 
PMTs are delivered by the manufacturer pre-tested and accompanied by a reference value for the (negative) high voltage, from now on H$_0$. This value is measured by the manufacturer in current mode operation and with a proprietary resistive base. The PMT HV needs to be properly tuned after coupling to the active base boards chosen by the KM3NeT Collaboration. The equalisation of the gain to the reference value is performed with a pulsed light source.

A sub-set of 600 PMTs was carefully investigated by the KM3NeT Collaboration. Using a resistive base and an ad-hoc setup, the HV of each PMT was tuned to obtain the reference gain of $3\times10^6$. The average value for the signal threshold corresponding to 0.3 photo-electrons, was determined to be 1113~mV. The 600 PMTs were then assembled with active base and re-tested. The time over threshold distribution was determined using the tuned HV and the average threshold values. The average value of  the time over threshold peak for single photo-electron signal was determined to be 26.4~ns.

The accurate, but time consuming, HV tuning procedure previously described cannot be applied for the mass test. Therefore, the KM3NeT Collaboration decided to use the reference time over threshold value of 26.4~ns with 1113~mV threshold as an indirect observable to equalise the PMT gains to  $3\times 10^6$ gain. The HV for all PMTs is optimized for a mass production in a way that, for a single photo-electron signal, the peak value of the time over threshold distribution occurs at 26.4~ns. 

\begin{figure}
\center
\includegraphics[width=0.45\textwidth]{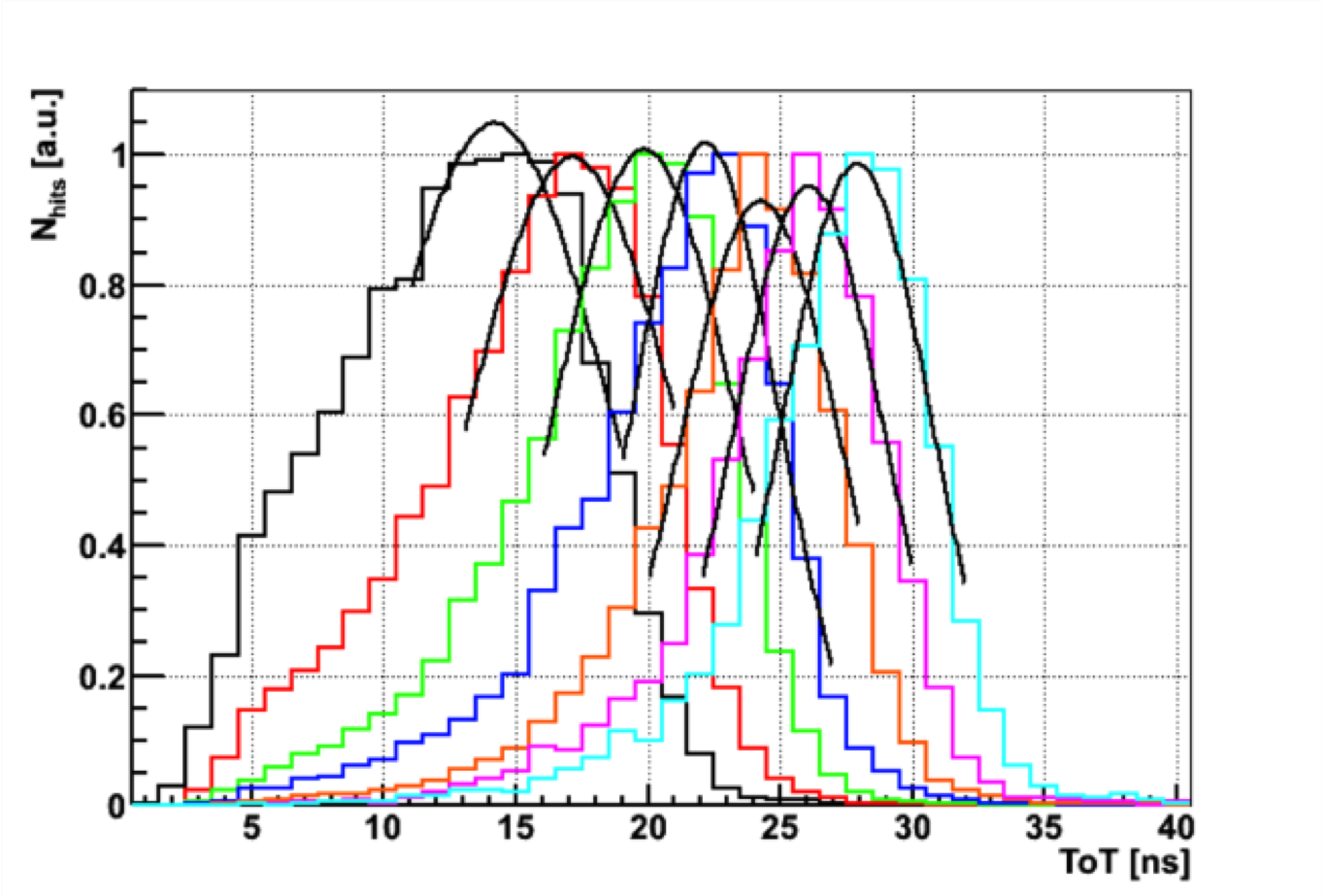}
\includegraphics[width=0.45\textwidth]{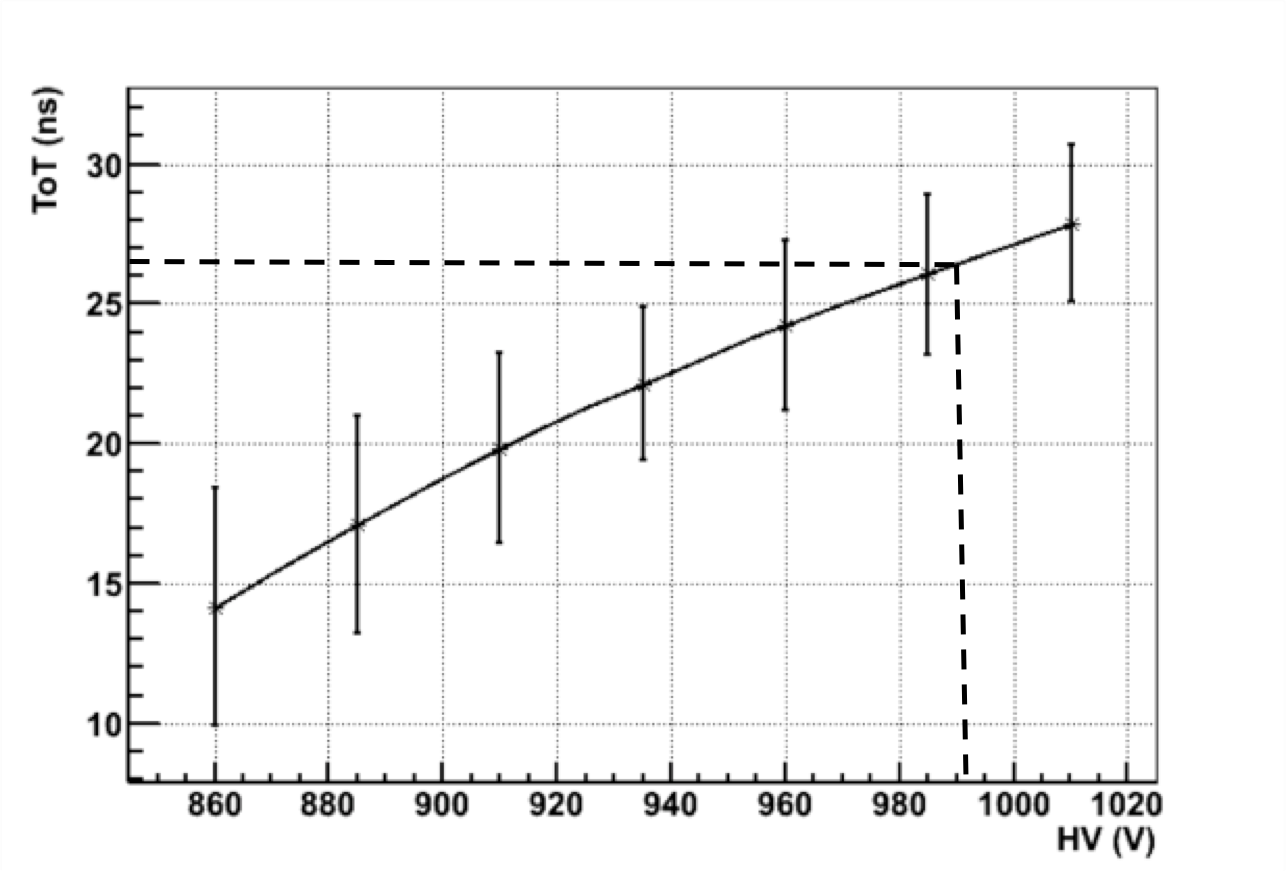}
\caption{Gaussian fit of time over threshold distributions (left) from HV$=H_0-100$~V (black) to $H_0 + 50$~V with a step of 25 V and parabolic fit of the time over threshold vs high voltage distribution (right). Dashed lines show 26.4~ns value and the chosen ``tuned'' high voltage value.}
\label{fig:hvtuning}
\end{figure}

The HV range ($[H_0-100\textrm{ V};\ H_0+50\textrm{ V}]$) is spanned in $25$~V steps. For each HV data noise impulses are acquired for about 5 minutes. The measured ToT data are collected and their distribution is fitted (Figure~\ref{fig:hvtuning}, left) with a Gaussian function in the range of [maximum bin - 4~ns; maximum bin + 4~ns] to determine the ToT peak. The measured peak is then plotted as a function of the PMT HV (7 points in Figure~\ref{fig:hvtuning}, right). 

The PMT gain grows almost linearly with HV in the operation range, so does the pulse area (pulse charge). The pulse width (ToT) dependence on the HV is more complicated. It was measured that, in the chosen HV range, ToT follows approximately a second order polynomial dependence on HV (Figure~\ref{fig:hvtuning}, right) with a negative coefficient for the quadratic term. So, data points are fitted with a parabola and the ``tuned HV'' value corresponding to a ToT of 26.4~ns is eventually determined (the smaller root value is used).

\subsubsection{Darkening and dark count measurements}
\begin{figure}
\center
\includegraphics[width=0.6\textwidth]{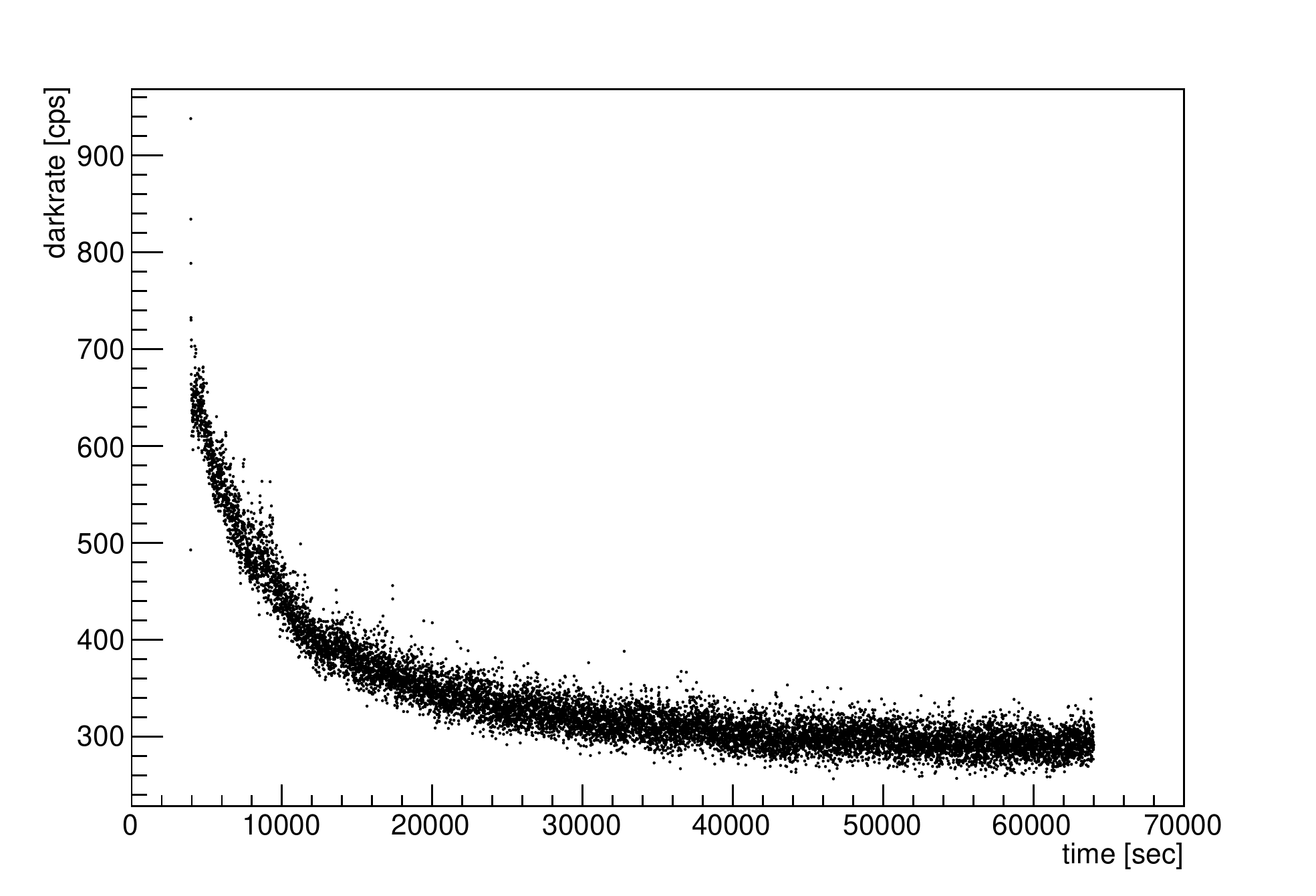}
\caption{An example of the dark rate measurements (in counts per second) during the darkening showing PMT recovery after exposure to the light.}
\label{fig:darkening}
\end{figure}

The ``tuned'' HV values are recorded in the database and set for each PMT. Dark count rates are then monitored for 9~hours. During this time PMTs recover from the initial exposure to the light and the dark count rate stabilizes (Figure~\ref{fig:darkening}). The final dark count rate is measured as the average value over the last 100~s of the run for each PMT. 

\subsubsection{Measurement of PMT time characteristics and of spurious pulses}
After the time needed for darkening a five-minute-long run is performed with PMTs illuminated by a laser whose trigger frequency is set to 20~kHz and the light output is set to operate PMTs in single photoelectron regime. This data is analyzed to estimate PMT timing performances and quantify spurious pulses.

{\bf Measurement of PMT timing performances.}
Time characteristics of PMTs are measured by detecting and analysing the so called first photon hits, i.e. pulses detected in the window [$T_0$, $T_0+200$~ns], where $T_0$ is the calculated arrival time of laser photons on the PMT surface. First hits must have no hits before them in the defined  time window. 

\begin{figure}
\center
\includegraphics[width=\textwidth]{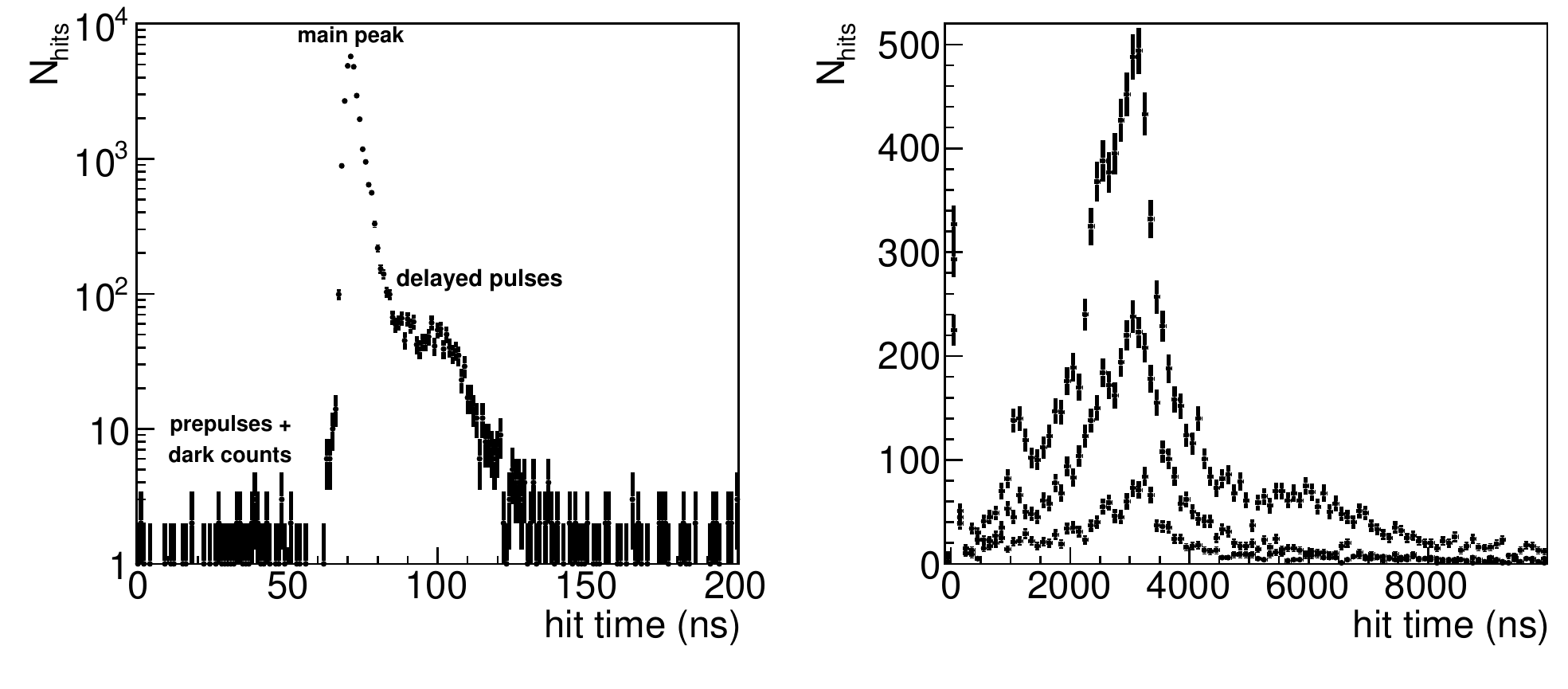}
\caption{Distribution of time of arrival of first hits for one PMT (left) and after pulses for three different PMTs (right). The exposure time is different between the both plots.}
\label{fig:first_hits}
\end{figure}

The time of arrival distribution of the first hits is shown in Figure~\ref{fig:first_hits} (left). The earliest high peak of the distribution corresponds to the PMT transit time, having already subtracted the travel time of photons in the optical path and the electronic time latency between the laser trigger signal and the laser optical output. These optical and electronics delays are fixed and measured during the DarkBox setting-up. 

The value of the transit time is determined as the centre of the bin with the maximum counts. The transit time spread is defined as the FWHM of the main peak. It is determined by counting the number of bins for which the content value is greater than half of the maximum bin content. Since the histogram binning is 1~ns, the error on FWHM is 1.4~ns.

Both transit time and its spread values are outcomes of this subphase.

{\bf Determination of PMT spurious pulses.}
The left-hand tail of the time distribution shown in Figure~\ref{fig:first_hits} (left) has a Gaussian shape, while for larger times the distribution is characterised by the presence of delayed pulses. The right slope of the main peak is softened by the delayed pulses due to inelastic scattering of the electrons on the dynodes~\cite{Lubs}. Additionally, the peak of delayed pulses, due to elastic scattering of the electrons on the dynodes, is observed at about 35~ns after the main peak. It was decided to calculate the percentage of delayed pulses as the ratio of the first hits in the window $[T_\textrm{peak} + 15.5\textrm{ ns}, T_\textrm{peak}  + 60.5\textrm{ ns}]$ over the number of the all first hits; $T_\textrm{peak}$ corresponds the centre of the maximum bin (transit time peak).

The percentage of prepulses, arriving before the main pulse (due to the through-passing photon and the following photoelectric effect on the first dynode), is also calculated. It is defined as the ratio of the hits in the window $[T_\textrm{peak} - 60.5\textrm{ ns}, T_\textrm{peak}-10.5\textrm{ ns}]$ over the number of the first hits. Their contribution is negligible for the studied PMTs.

Afterpulses are defined as hits with a first hit before them in a time window of 10~$\mu$s. The time distribution of afterpulses for several PMTs is shown in Figure~\ref{fig:first_hits} (right). The plateau of the distribution is due to the dark noise and it is reached at $\sim10$~$\mu$s for most of the PMTs.

The used front-end electronics does not allow a good separation for the consecutive hits with a time difference $\lesssim10$~ns~--- signal discrimination with 0.3~p.e. threshold concatenates both impulses to a single long one. Fast afterpulses (Type I, due to the backscattered secondary electrons or photo-production at the dynodes) arrive $\le20$~ns after the first hits, so their quantification was not possible with current setup.

The percentage of slow afterpulses (Type II, due to ionization of the gas molecules in the tube) is determined as the ratio of the afterpulses in the time window $[T_\textrm{peak} + 100.5\textrm{ ns}, T_\textrm{peak} + 10\textrm{ }\mu\textrm{s}]$ over the number of the first hits.

The percentages of spurious pulses of all types are eventually corrected taking into account the dark noise hits that contaminate both first hits and spurious pulse distributions.

\begin{figure}
\center
\includegraphics[width=\textwidth]{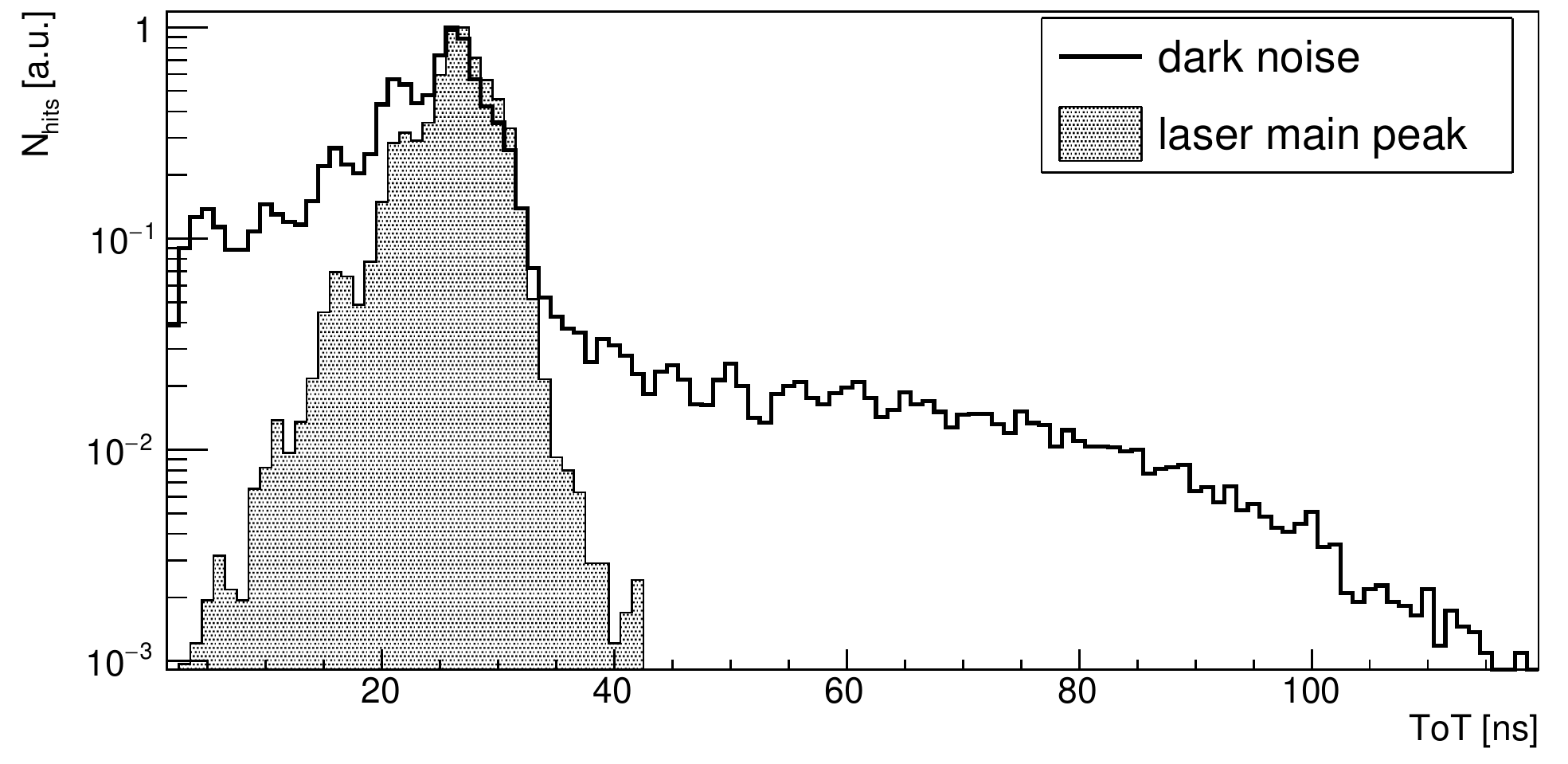}
\caption{Time over threshold distributions for dark noise hits and hits in the main laser peak.}
\label{fig:tot}
\end{figure}

{\bf Check of time over threshold distribution.}
During the run with laser, the ToT distribution of the first hits arriving in the time window around transit time peak with a width of the transit time spread is collected.
This distribution is shown in Figure~\ref{fig:tot} and compared to the one determined using dark count hits.
The shape of the distribution obtained using laser photons is better fitted with a Gaussian with respect to the one obtained with dark noise hits: this feature confirms that the laser is correctly operated in single photo-electron mode. The average ToT is then measured fitting the distribution with a Gaussian function in the window $[\textrm{ToT}_\textrm{peak}-4.5\textrm{ ns}; \textrm{ToT}_\textrm{peak} + 4.5\textrm{ ns}]$, where $\textrm{ToT}_\textrm{peak}$ corresponds to the centre of the ToT distribution bin with maximum counts.

\section{System performance}
\label{s:qualification}
In order to validate the DarkBox instrument we verified that all 62 sockets and their electronic channels are identical and, thus, provide the same result during the various steps of the PMT calibration procedure.

\begin{figure}
\center
\includegraphics[height=0.43\textwidth]{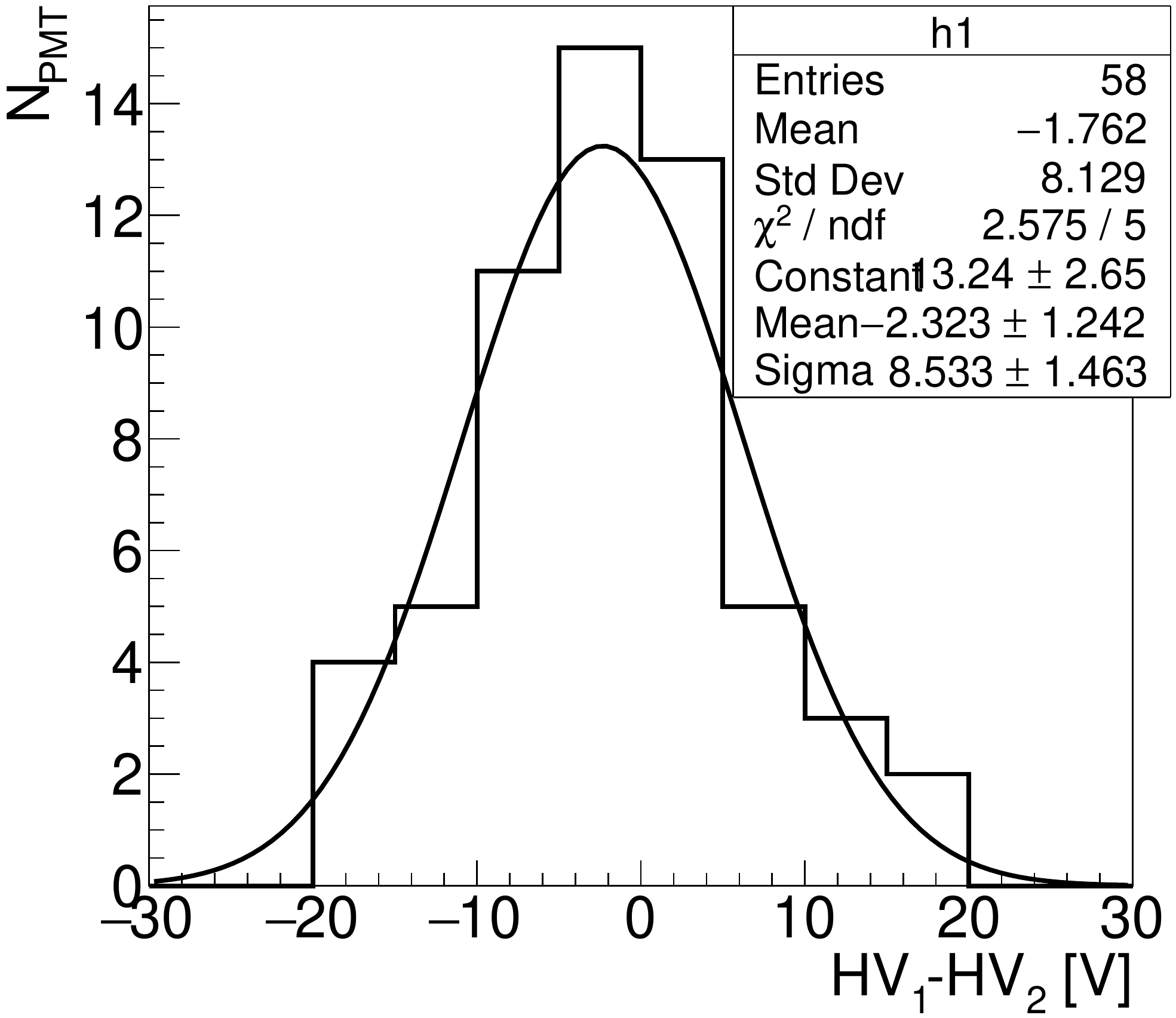}
\includegraphics[height=0.43\textwidth]{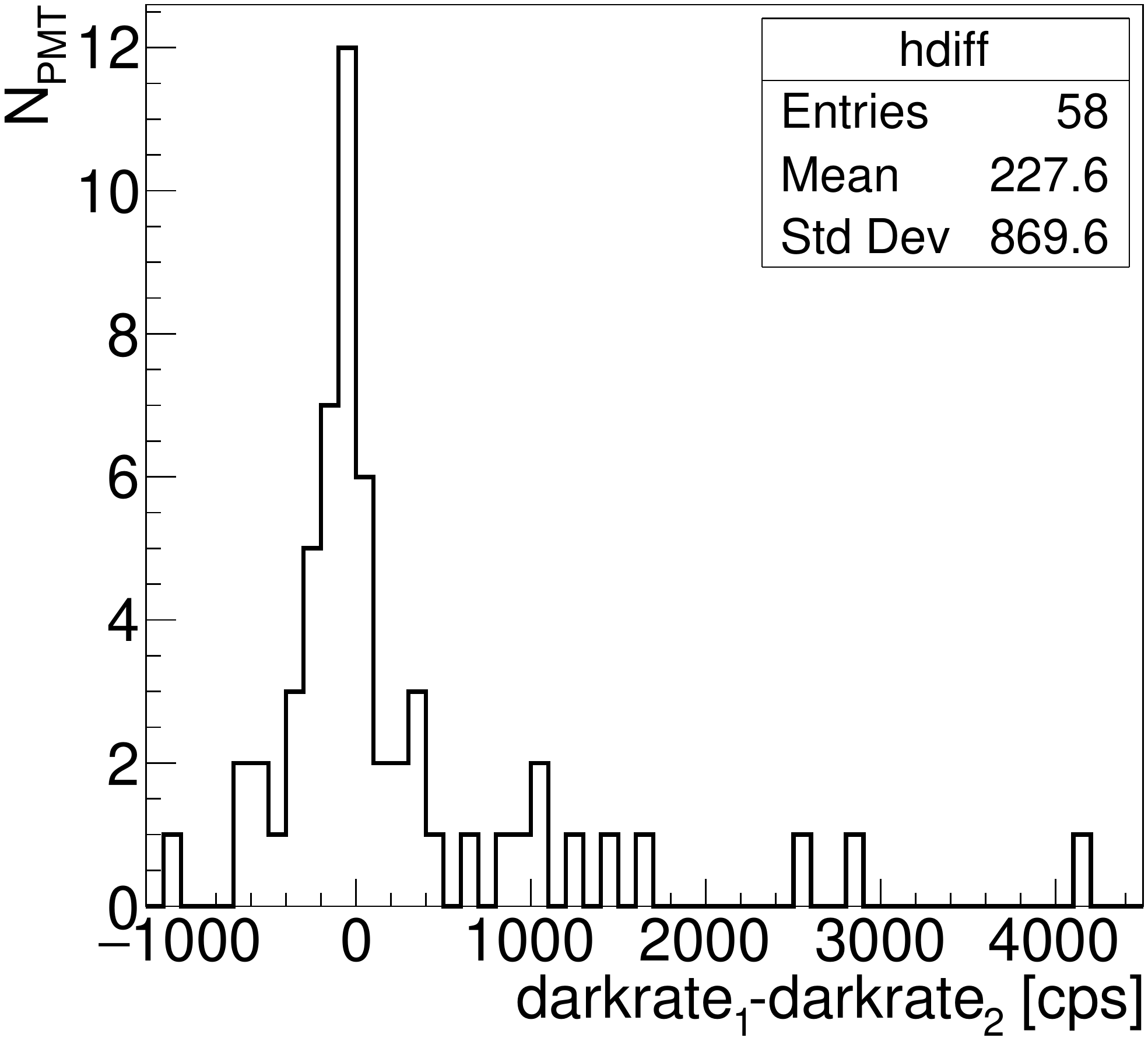}
\caption{PMT by PMT difference of the tuned HV values (left) and the dark rates in counts per second (right) between two measurements with the same set of the PMTs placed randomly in the DarkBox sockets.}
\label{fig:hv_tuning_test}
\end{figure}

{\bf Sockets uniformity: HV equalising.} The HV tuning procedure was carried out on a batch of 62 PMTs. The same procedure was then repeated after randomly exchanging the PMT positions in the sockets.  The distribution of the difference between the optimized HV  determined during the two runs is shown in Figure~\ref{fig:hv_tuning_test} (left). 

This validation procedure showed that for 58 PMTs the HV is measured with a precision of about 9 V. For 3 PMTs data corruption caused failure of HV tuning; in addition there is only one outlier with 80~V difference between two tunings (not shown in Figure~\ref{fig:hv_tuning_test}). These problems were induced by an abnormal ToT distribution caused by electrical signal reflections due to not perfect cable connection. Re-testing such PMTs in different sockets solves the problem.

{\bf Sockets uniformity: dark rates.} Additionally, we tested the capability of the DarkBox to measure the dark counts independently of the PMT socket position. 
As well known from the literature, the response of negatively fed photomultipliers (e.g. the KM3NeT ones) is rather unstable being strongly dependent on the material surrounding the photocathode and the dynode glass envelope. This feature may lead to very different dark count rates depending on the mechanical coupling of the PMT with the support structure. 
The dark count rate differences measured in the two tests (that is before and after redistributing PMTs randomly in the trays) are shown in the right panel of Figure \ref{fig:hv_tuning_test} (right). Only for 5 out of 58 PMTs the measured dark count rate considerably changed. We could then conclude that the reproducibility of the dark count measurement in the DarkBox  is about 90\%. In order to prove that, we swapped again in a third test the positions of the 5 PMTs that showed extremely high dark count rate in the second test.  After this operation these 5 PMTs produced the dark count rates similar to the rates in the first test.

\begin{figure}
\center
\includegraphics[width=.9\textwidth]{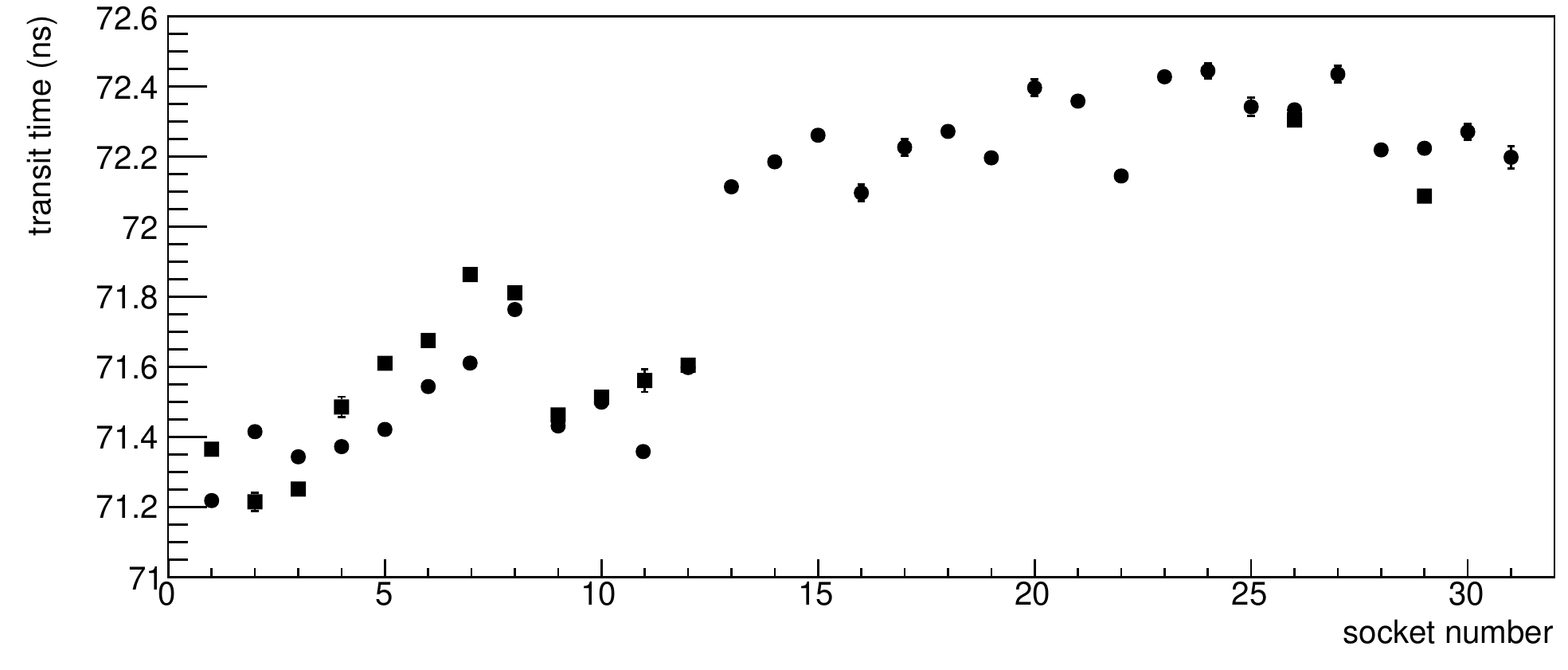}
\caption{Measured Transit Time of the reference PMT installed on different sockets of the first Dark Box tray (circles) and some sockets of the second tray (squares). Systematic errors of about 100~ps are not shown. Observed spread is due to the different copper paths of the KM3NeT front end electronics (see text).}
\label{fig:tts_diff_chan}
\end{figure}

{\bf Sockets uniformity: timing.} To verify the equivalence of time measurement performances for the different sockets, a reference PMT was installed and tested over all DarkBox sockets. The measured transit time spread was equal to 4~ns for all the sockets. The transit time distribution is shown in Figure~\ref{fig:tts_diff_chan}. It can be seen that the PMTs placed in the first 12 sockets have shorter transit time with respect to the remaining 19 sockets (780~ps in average).
This is due to a peculiar behavior of  the KM3NeT electronics boards: PMTs in these 12 and 19 sockets are connected to the two different octopus boards (small and large, correspondingly) and their signal routing is different (signal paths on the large octopus board are longer). The differences of transit time values inside each group (with 12 and 19 PMTs) is below 200~ps, as the signal paths have similar lengths inside each octopus board. The difference between the two trays, when the PMT is placed in the socket with the same number, is about 100~ps, as both control boards acquiring data from the trays are synchronized. The differences 200~ps and 100~ps are negligible for KM3NeT application.

{\bf Sockets uniformity: spurious pulses.} The benchmark PMT spurious pulses percentage is reported in Table~\ref{t:spurious_bench} for all sockets.  The differences are small and within statistical errors.

\begin{table}
\caption{Spurious pulse ratios for the benchmark PMT rotated over all DarkBox sockets.}
\centering
\label{t:spurious_bench}
\begin{tabular}{l c c c}
Type & Min & Max & Avg\\
\hline 
Prepulses & 0.0034\% & 0.023\% & 0.012\% \\
Delayed & 2.99\% & 3.24\% & 3.15\% \\
After & 3.037\% & 3.67\% & 3.22\%\\
\end{tabular}
\end{table}

{\bf System performance summary.} The validation procedure described above showed that all PMT sockets in the DarkBox are equivalent for the purpose of the application.

In addition it was shown that a single PMT test in the DarkBox has a percentage of success close to 90\%. This means that 10\% of PMTs have to be tested twice, which is acceptable for mass production.

\section{Conclusions}
The paper describes a novel instrument designed for massive test and calibration of small size PMTs. The instrument permits to measure PMT dark rate, equalise gain, determine spurious pulses and timing characteristics (transit time and transit time spread).
The setup has 62 sockets that allow massive test and characterisation of 3-inch PMTs. The use of interchangeable trays to host PMTs and ease of connections allows to complete calibration in about 10 hours (9 hours are used for PMT conditioning in darkness). The equivalence between different PMT sockets of the DarkBox has been demonstrated using benchmark PMTs. 
 
The mechanical and laser signal distribution systems are totally general-purpose. Front-end electronics and data acquisition system were customised in order to test and calibrate PMTs used in KM3NeT. The DarkBox system and the developed calibration procedure is, however,  extremely flexible. Therefore, it can be adapted to other PMT types or data acquisition electronics with minimal changes.

In conclusion, we developed a powerful tool for any experiment requiring massive test and calibration of PMTs, which can be adapted or reproduced following the design and procedures shown in the paper.

\acknowledgments
We thank the KM3NeT Collaboration for providing benchmark PMT assemblies, front end and data acquisiion systems, and the Technical Project Manager of KM3NeT, Marco Circella of INFN Bari, for his advice on the optimization of the design.

\end{document}